\documentclass[12pt]{article}
\pdfoutput=1
\usepackage{epsfig}
\usepackage{graphicx}
\usepackage{cite}
\usepackage{amsfonts}
\usepackage{amssymb}
\usepackage{bm}
\usepackage{latexsym}
\usepackage{amsmath}
\numberwithin{equation}{section}
\def\ee{\end{equation}}
\def\ba{\begin{eqnarray}}
\def\ea{\end{eqnarray}}

\def\bq{\begin{quote}}
\def\eq{\end{quote}}

 at 10truept

\newcommand{\beq}{\begin{equation}}
\newcommand{\eeq}{\end{equation}}
\newcommand{\beqa}{\begin{eqnarray}}
\newcommand{\eeqa}{\end{eqnarray}}
\newcommand{\bea}{\begin{eqnarray}}
\newcommand{\eea}{\end{eqnarray}}

\def\lesssim{~\mbox{\raisebox{-.6ex}{$\stackrel{<}{\sim}$}}~}

\def\ltap{\ \raise.3ex\hbox{$<$\kern-.75em\lower1ex\hbox{$\sim$}}\ }
\def\gtap{\ \raise.3ex\hbox{$>$\kern-.75em\lower1ex\hbox{$\sim$}}\ }
\def\gl{\ \raise.5ex\hbox{$>$}\kern-.8em\lower.5ex\hbox{$<$}\ }
\def\roughly#1{\raise.3ex\hbox{$#1$\kern-.75em\lower1ex\hbox{$\sim$}}}

\begin{document}

\thispagestyle{empty}
\begin{titlepage}
\nopagebreak

\title{
Astronomical tests for quantum black hole structure} 
 
 %\begin{center}
%\end{center}

\vfill
\author{Steven B. Giddings\footnote{giddings@ucsb.edu}\ 
}
\date{ }

%\end{center}

\maketitle

\vskip 0.5cm

\begin{center}
 Department of Physics, 
University of California, Santa Barbara, CA 93106

\begin{abstract}
Quantum modifications  to black holes on scales comparable to the horizon size, or even more radical physics, are apparently needed to reconcile the existence of black holes with the principles of quantum mechanics. This piece gives an overview of some possible observational tests for such departures from a classical description of black holes, via gravitational wave detection and very long baseline interferometry.   (Invited comment for {\sl Nature Astronomy}.)
\end{abstract}

\end{center}
\vfill
 \vskip.4in
 
\noindent
%CERN-PH-TH/2010-095\hfill \\  
\vfill
\end{titlepage}

\setcounter{equation}{0} \setcounter{footnote}{0}

Black holes present a profound challenge to our current foundations of physics, and an exciting  era of astronomy is just opening where    gravitational wave observation, represented by the recent LIGO detections\cite{Abbott:2016blz}, and very long baseline interferometry (VLBI), which is on the verge of resolving the immediate environs of super-massive black holes\cite{Doeleman:2009te}, may provide important hints about the new principles of physics  needed.

While the near-horizon region of a black hole is intrinsically interesting and has not been directly probed before, the conventional expectation has been that physics here, for astrophysical black holes, will be governed by classical general relativity (GR), and so the new observations should simply confirm GR's validity.  After all, the spacetime curvature near such a horizon -- a basic measure of the local gravity -- would be tiny as compared to the values expected to be associated with any quantum behavior of spacetime.

However, Hawking's discovery that black holes evaporate\cite{Hawking:1974sw} has shaken this view to the core, with many quantum physicists now concluding that this result implies significant modifications to spacetime near black holes, and perhaps even further away.  In fact, an attempt to give a logically consistent description of the evolution of black holes has lead to a foundational crisis in physics, which is apparently difficult to resolve without modifications to conventional physics not just at very short distance scales, as is expected in quantum gravity, but on macroscopic scales comparable to horizon radii of large black holes.

The problem arises when considering information that falls into a black hole.    In our most basic framework for currently established physics, quantum field theory, information cannot propagate faster than light, so then cannot escape.  Therefore if a black hole evaporates away and disappears, as Hawking predicted, it  destroys the information that fell in.  This violates a bedrock principle of quantum mechanics, unitarity.  Physicists have struggled with this issue for forty years\cite{Hawking:1974sw}, without finding a logically consistent resolution respecting current principles of physics.  This problem appears to represent a profound conflict between the principles of relativity, the principles of quantum mechanics, and the principle of locality, which are the foundational principles of quantum field theory -- our most fundamental current description of reality.  

Apparently something must give.  Attempts to modify quantum mechanics have led to nonsensical conclusions, so many now take it as inviolate.  But, to preserve information, and save quantum mechanics, apparently there must be new physical effects operating near black holes, capable of conveying information from their deep interiors to the vicinity of their horizons.  Finding a consistent description of such black-hole scale effects has been a driving theme in recent theory.  

Escape of information from a black hole appears to require a modification of the principle of locality, which in quantum field theory is stated as the prohibition of faster-than-light information transfer.  This seemingly crazy conclusion first began to be taken seriously twenty five years ago\cite{BHMR}.  More recently, in an attempt to contain the damage, and motivated by considerations of string theory, it was proposed\cite{Almheiri:2012rt} that such new effects stop sharply within a microscopic distance of the horizon -- but that has led to the wild conclusion that an energetic curtain of particles, called a ``firewall," {\it replaces} the horizon.  Another string-inspired possibility is that the higher-dimensional geometry of string theory provides extra microscopic structure outside a would-be horizon, the so-called ``fuzzball" scenario\cite{Mathur:2008nj}.  Finally, \cite{NVNL} has advocated the possibility of a ``softer" violation of locality.  This could arise from quantum interactions between the internal state of the black hole and the quantum fields near the black hole, needed to transfer information out; since these would appear to be nonlocal, they may well not arise from a quantum field theory description of physics, and so may owe their existence to some more basic description of quantum spacetime.   {\it Each} of these scenarios modifies standard GR on scales comparable to the event horizon size $R_H$ -- a realm we are now observationally investigating.

A first characterization of such possible new physics is to describe its relevant scales.  If a black hole has some such new structure -- corresponding to a departure from the vacuum behavior expected from GR near the black hole -- it may extend a distance $R_{\alpha}$ outside the horizon and spatially vary on a scale $L\lesssim R_{\alpha}$.  The scale $L$ then would characterize the ``hardness" of  interactions, {\it e.g.} of matter, with  that structure, since as is seen for example by the uncertainty principle, interactions varying on scales $L$ typically yield momentum transfers $\sim \hbar/L$.  These scales -- and the strength of the new effects -- are important in describing their possible observational consequences, since $R_{\alpha}$ indicates how far these modifications reach, and, for example, smaller $L$ -- or harder structure -- means larger possible departures from the matter and light motion predicted by GR.  Specifically, the expected VLBI images, {\it e.g.} from the Event Horizon Telescope (EHT), are generated by light emitted from accreting matter within a region of size $\sim R_H$ surrounding the horizon, and so departures from a GR description of spacetime in a region given by $R_{\alpha}\sim R_H$ could potentially distort these images.  Likewise, in black hole collisions, part of the gravitational-wave  signal is emitted in the plunge/merger phase where black holes closely approach and coalesce, so LIGO/VIRGO is also potentially sensitive to deviations at these distances.

Firewalls, with an expected $R_{\alpha}\sim L$ given by a microscopic scale ({\it e.g} the Planck length) are probably hard to test with such observations, unless there are reflection effects (see {\it e.g.} \cite{CFP, CHMPP, Refl, PrKh, NSTT}) from the firewall.  For fuzzballs, one expects $L$ to be microscopic, but there is no clear prediction for $R_{\alpha}$.  It now appears possible to begin to rule out such hard structure if it extends to $R_{\alpha}\sim R_H$, due to the closeness of the LIGO signal GW150914 to the GR prediction\cite{Abbott:2016blz,Giddings:2016tla}; neutron star simulations demonstrate that hard structure has noticeable impact on gravitational wave signals.  An important question is the possible testability of the ``soft quantum structure" of \cite{NVNL}, which may have scales $R_{\alpha}\sim L \sim R_H$.  Its potential impact on LIGO signals requires more complete knowledge or modeling of its evolution in the region where gravity becomes nonlinear, near a black hole.
There are not, yet, precise predictions of the deviations in gravitational wave signals that {\it any} of the candidates for quantum BH structure would produce. Theoretical development is clearly needed to improve the situation, including potentially via phenomenological models.  Since we expect many  more BH coalescence events\cite{LIGOrate}, it may be that statistical means can be used to search for and constrain deviations from GR with increasing statistical power.  

VLBI observations may offer even simpler tests of some of these scenarios.  Since these measure electromagnetic radiation emitted by infalling matter, they are sensitive to perturbations in the BH geometry away from that predicted by GR, without requiring the full nonlinear dynamical evolution law that governs these perturbations; photons just give us a ``picture" of the geometry, unlike gravitational wave signals whose interpretation depends on the nonlinear gravitational evolution.  For that reason there is great interest in investigating the possible impact of quantum modifications to BH structure on the images that are expected from EHT.  GR predicts images with a particular structure, and specifically a ``shadow" of the black hole with a bright ``photon ring" just outside, resulting from gravitational lensing around the black hole.  
For example, in the case of soft quantum structure of \cite{NVNL}, the necessary transfer of information may be accomplished by fluctuations with large {\it magnitude}, and simple models of such fluctuations can have significant impact on predicted images\cite{Giddings:2016btb}, including distortion of the photon ring and shadow, and the possibility of observable time dependence of the image.  An open question is whether other versions of this soft scenario permit smaller, less-visible fluctuations, as newer work is suggesting\cite{NVU}.  Observation is an important guide.  The first EHT images with good event-horizon scale resolution are eagerly awaited, and it will be very interesting to see if there are any deviations from the GR-predicted images, or how strongly deviations can be bounded.  If deviations are observed, they could be a a signature of the strong, soft quantum structure just described.

In summary, from the point of view of fundamental physics there are good reasons to expect that BH structure is modified on scales comparable to the horizon size,  but there is still a lot of debate in the theoretical community about the precise form of the modified structure.  Observational constraints are therefore  valuable to examine the predictions of GR, and investigate the possibility of departures from it.
 With the advent of gravitational wave astronomy and the increasing power of VLBI, astronomers have dual opportunities to look for possible new effects, indicating modifications to fundamental physics, or to tighten constraints that paint theorists more tightly into a corner as they try to resolve the quantum mysteries of black holes.

\vskip.3in
\noindent{\bf Acknowledgements}

This work  was supported in part by the Department of Energy under Contract No. DE-SC0011702, and by Foundational Questions Institute (fqxi.org) 
grant FQXi-RFP-1507.

\end{document}